\documentstyle[12pt]{article}

\begin{document}
\def\bib#1{[{\ref{#1}}]}
\def\at{\tilde{a}}

\begin{titlepage}
         \title{Classical Electrodynamics of a Particle \\ 
   with Maximal Acceleration Corrections\footnote{Research supported 
by MURST fund 40\% and 60\%, DPR 382/80}}

\author{A. Feoli$^{a}$\thanks{{\it{E-mail}}: Feoli@vaxsa.csied.unisa.it}
~~G. Lambiase$^{a}$\thanks{{\it{E-mail}}: lambiase@vaxsa.csied.unisa.it}
~~G. Papini$^{b}$\thanks{{\it{E-mail}}: papini@cas.uregina.ca}
~~G. Scarpetta$^{a,c}$ \\
{\em $^a$Dipartimento di Fisica Teorica e s.m.s.a.}\\ 
{\em  Universit\`a di Salerno, 84081 Baronissi (SA), Italy}\\
{\em  $^a$Istituto Nazionale di Fisica Nucleare, Sezione di Napoli}\\
{\em $^b$Department of Physics, University of Regina,} \\
{\em Regina, Sask. S4S 0A2, Canada}\\
{\em $^c$International Institute for Advanced Scientific Studies} \\ 
{\em Vietri sul Mare (SA), Italy}}
              \date{\empty}
              \maketitle

              \begin{abstract}
We calculate the first order maximal acceleration corrections to the
classical electrodynamics of a particle in external electromagnetic fields.
These include additional dissipation terms,
the presence of a critical
electric field, a correction to the cyclotron frequency of an electron 
in a constant magnetic field and the power radiated by the particle.
The electric effects are sizeble at the fields that are considered 
attainable with ultrashort $TW$ laser pulses on plasmas.  

	      \end{abstract}

\thispagestyle{empty}
\vspace{20. mm}
              \vfill
	      \end{titlepage}

\section{Introduction}
\setcounter{equation}{0}

In an attempt to incorporate quantum aspects in the geometry of space-time, 
Caianiello wrote a series of papers about a theory called "Quantum
Geometry" \cite{qg}, where the position and momentum 
operators are represented as
covariant derivatives and the quantization is interpreted as curvature of phase
space. A novel feature of the model is the existence of a maximal acceleration  
\cite{am}.
Limiting values for the acceleration were also derived
by several authors on different grounds and applied to many branches 
of physics such as string
theory, cosmology, quantum field theory, 
black hole physics, etc. \cite{prove}.

The model proposed by Caianiello and his coworkers to include the effects of 
a maximal acceleration in a particle dynamics consisted in enlarging 
the space-time manifold
to an eight-dimensional space-time tangent bundle TM.
In this way the invariant line element becomes
\begin{equation}
d\tilde{\tau}^{2}=g_{AB}dX^{A}dX^{B},\quad A,B=1,\ldots,8
\end{equation}
where the coordinates of TM are
\begin{equation}
X^{A}=\left(x^{\mu};{\frac{c}{A}}{\frac{dx^{\mu}}{d\tau}}\right),\quad
{\mu=1,\ldots,4}
\end{equation}
$A=mc^3/\hbar$, and
\begin{equation}
g_{AB}=g_{\mu\nu}\otimes g_{\mu\nu}\,{.} 
\end{equation}
The metric (1.1) can be written as
\begin{equation}
d \tilde {\tau}^{2}=g_{\mu\nu}(dx^{\mu}dx^{\nu}+\frac{c^2}{A^2} d\dot{x}^{\mu}
d \dot{x}^{\nu})\,{,}
\end{equation}
and an embedding procedure can be developed \cite{em} in order 
to find the effective space-time
geometry in which a particle moves when the constraint of a maximal
acceleration is present.
In fact, if we find the parametric equations that relate the velocity field 
$\dot{x}^{\mu}$ to the first four coordinates $x^{\mu}$, we can calculate the
effective four dimensional metric $\tilde {g}_{\mu\nu}$ on the hypersurface
locally embedded in TM.

In Section 2 we derive the quantum corrections to the metric for a charged 
particle moving in an electromagnetic field and the corresponding first 
order corrections to the equations of motion. 
In Section 3 we analyze some 
consequences of the corrections for particular instances. 
We derive the radiated power in Section 4. Section 5 contains a discussion.

\section{Effective Geometry}
\setcounter{equation}{0}

Consider the Minkowski metric $\eta_{\mu\nu}$ (of signature -2) as the background metric in Eq. (1.4). 
A particle of mass $m$ and charge $q$ moves in this background under the 
action of an externally applied 
electromagnetic field. 
The  classical equation of motion of the particle is 
\begin{equation}  
d\dot{x}^{\mu}=-\frac{q}{mc}F^{\mu}_{\,\,\,\nu}d{x}^{\nu}
\end{equation}
where $F_{\mu\nu}$ is the classical electromagnetic field tensor. Eq. (2.1) may
be taken as the first order
approximation to the real velocity field of the particle.
If we substitute Eq. (2.1) into (1.4) we can calculate the correction of order
 $A^{-2}$ to the
classical background metric. We can iterate this procedure, by
calculating the new velocity field and substituting it again into 
the metric to
obtain the correction $A^{-4}$, and so on.
However the value of the maximal acceleration is very high and we can neglect
the $O(1/A^{4})$ terms. This leads to the new metric 
\begin{equation}
d\tilde {\tau}^{2}\simeq\left(\eta_{\mu\nu}-\frac{q^2}{A^{2}m^{2}}F_{\mu\lambda}
F^{\lambda}_{\,\,\,\nu}\right)dx^{\mu}dx^{\nu}
\equiv \tilde{g}_{\mu\nu}dx^{\mu}dx^{\nu}\,{.}
\end{equation}
The effective geometry is curved by the acceleration due to the interaction of the charged particle
with the electromagnetic field and this curvature affects the motion of the 
particle itself.

The modified equation of motion of the charged particle may be 
obtained from the action 
\begin{equation}
S= \int{(-mc^{2}\sqrt{\tilde{g}_{\mu\nu} \dot{x}^{\mu} \dot{x}^{\nu}}
+q A_{\mu} \dot{x}^{\mu})d\tau}
\end{equation}
and leads to the equations
\begin{equation}
\ddot{x}^{\sigma}+
{\Gamma}^{\sigma}_{\,\,\,\alpha\beta}\dot{x}^{\alpha}\dot{x}^{\beta}
+\frac{q}{mc}\tilde{g}^{\sigma\rho}F_{\rho\gamma}\dot{x}^{\gamma}=0\,{,}
\end{equation}
where  
\begin{equation}
\Gamma^{\sigma}_{\,\,\,\alpha\beta}=\frac{q^2}{2A^{2}m^{2}}\tilde{g}^{\sigma\nu}
[F^{\lambda}_{\,\,\,\nu}(F_{\lambda\alpha,\beta}+F_{\lambda\beta,\alpha})+
F^{\lambda}_{\,\,\,\alpha}F_{\beta\nu,\lambda}
+F^{\lambda}_{\,\,\,\beta}F_{\alpha\nu,\lambda}]\,
\end{equation}
is the connection,
and the dot means derivation with respect to $\tilde\tau$.

The contravariant components of the  
metric tensor $\tilde{g}^{\sigma\nu}$ then follow from the equation 
$\tilde{g}^{\sigma\nu}
\tilde{g}_{\nu\rho}=\delta^{\sigma}_{\rho}$ in the approximation $\frac{q^2}
{A^{2}m^{2}} F^{\sigma}_{\, \, \lambda}F^{\lambda\nu}\ll\eta^{\sigma\nu}$ 
 and are
\begin{equation}
\tilde{g}^{\sigma\nu}\simeq\eta^{\sigma\nu}+\frac{q^2}{A^{2}m^{2}}
F^{\sigma}_{\,\,\,\lambda} F^{\lambda\nu}\,{.}
\end{equation}
Hereafter the raising and lowering of the indices is made with the
tensor $\eta_{\mu\nu}$.
Obviously the corrections disappear in the limit $A\to\infty$ and 
we find the classical equation of motion.

To analyze the physical consequences of this correction, it is useful to 
write the equation of motion (2.4) in terms of the four-momentum $p^{\mu}=
mc\dot{x}^{\mu}$ and the four-velocity $ dx^{\mu}/d\tilde\tau = u^{\mu}
= (c dt/d\tilde\tau, \vec v dt/d\tilde\tau)$
\begin{equation}
\frac{dp^{\mu}}{d\tilde\tau}=\frac{q}{c}\tilde{g}^{\mu\nu}F_{\nu\rho}u^{\rho}-
\frac{q^2}{2A^{2}m}\tilde{g}^{\mu\nu}(F^{\lambda}_{\,\,\,\nu}
F_{\lambda\alpha,\beta}+
F^{\lambda}_{\,\,\,\alpha}F_{\beta\nu,\lambda})u^{\alpha}u^{\beta}\,{.}
\end{equation}

\section{First Order Corrections}
\setcounter{equation}{0}

The particle embedded in the external field experiences a curved
manifold even if the background metric is flat. 
This curvature affects the motion of the particle and the rate
of change of its energy $\varepsilon \equiv p^{0}$ in time.
 We analyze first the simple case when only an electric field
$\vec{E}$ is present.
 We obtain from (2.7)
\begin{equation}
\frac{d\varepsilon}{dt}=q{\vec{E}\cdot\vec{v}}\left(1+
\frac{q^{2}|\vec{E}|^{2}}{A^{2}m^{2}}\right)
+\frac{q^{2}c^{3}}{A^{2}m{\frac{d\tilde\tau}{dt}}}
\left[\frac{1}{2}\frac{d|\vec{E}|^{2}}{dt}+\frac{1}{c^2}
\left(\vec{v}\cdot\frac{\partial
\vec{E}}{\partial{t}}\right)\vec{E}\cdot\vec{v}+v^{j}E_{i}\partial_{i}
{E_{j}}\right]\,{.}
\end{equation}
The difference of (3.1) from the usual classical equation becomes evident
when $\vec{E}\cdot\vec{v}=0$. In this case we find
\begin{equation}
\frac{d\varepsilon}{dt}=
\frac{q^{2}c^{3}}{2A^{2}m{\frac{d\tilde\tau}{dt}}}\left[\frac{d|\vec{E}|^{2}}
{dt}+2v^{j}E_{i}\partial_{i}E_{j}\right]\,{,}
\end{equation}
which is an additional dissipation term.
On the other hand the orthogonality condition of $\vec{E}$ and
$\vec{v}$ cannot be imposed in general because of the perturbing effect of
the acceleration on the motion itself.

The quantity ${\frac{d\tilde\tau}{dt}}$ in Eq. (3.1) is defined by
\begin{equation}
{\frac{d\tilde\tau}{dt}}=c\sqrt{1-\frac{v^2}{c^2}-\frac{q^{2}|\vec{E}|^{2}}{A^{2}m^{2}}}\,{.}
\end{equation}
Since ${\frac{d\tilde\tau}{dt}}$ must be real, we have
\begin{equation}
|\vec{E}|<\frac{mA}{|q|}\sqrt{1-\frac{v^{2}}{c^2}}\,{,}
\end{equation}
i.e.,the intensity $|\vec{E}|$ is limited by the critical electric
field
\begin{equation}
E_{c}=\frac{mA}{|q|}\sqrt{1-\frac{v^2}{c^2}}\,{.}
\end{equation}
In the particle's rest frame $E_c$ becomes
\begin{equation}
E_{c}=\frac{mA}{|q|}\,{.}
\end{equation}
If the particle is an electron, we find
$E_{c}(m_{e^{-}})\simeq 2.1\cdot{10^{16}}N/C$
when the maximal acceleration is mass-dependent, and
$E_{c} \simeq 5.6\cdot{10^{39}}N/C$
when $A$ refers to the Planck mass. Similarly for the proton we find:
$E_{c}(m_{p^{+}})\simeq 10^{24}N/C$
and
$E_{c}\simeq 10^{43} N/C$ respectively.

It is interesting to observe that a
similar critical value for the electric field was found for the open bosonic 
string propagating in an external electromagnetic field. In this case 
$E<T/|q_{a}|\,\, (a=1,2)$ where $T$ is the string tension and $q_{a}$ are the 
charges at the ends of the string \cite{bnb}. This condition is satisfied 
automatically when the background electromagnetic field is described by the 
Born-Infeld Lagrangian rather then by the Maxwell Lagrangian \cite{BN}.
It is therefore possible to expect a connection between the Born-Infeld 
and the Caianiello Lagrangians
\cite{pre}. The relation between acceleration and $E_c$
was also analyzed by Gasperini \cite{ga} 

It is also instructive to calculate Eq. (2.7) explicitly in the case of
a particle with charge $q=-e$, moving in
a constant electromagnetic field of components $\vec{E}=(E_{x},E_{y},0)$
and $\vec{B}=(0,0,B)$. We find
\begin{equation}
\frac{d\varepsilon}{dt}=-e(E_{x}\dot{x}+E_{y}\dot{y})+\frac{e^{3}}{A^{2}
m^{2}}(E_{x}\dot{x}+E_{y}\dot{y})(|\vec{E}|^{2}-B^{2})\,{.}
\end{equation}
In this case all dissipation terms 
due to the maximal acceleration disappear only if $ \vec{E} = 0$.

Finally, we consider the problem of charged particles moving in a constant
magnetic field of components $\vec{B}=(0,0,B)$. 
From equation (2.7), the equations of motion for a circular path are
\begin{equation}
\left\{ \begin{array}{ll}x(t)=r_{0}\sin\omega{t} 
\\y(t)=r_{0}\cos\omega{t}\end
{array}\right.\,{,}
\end{equation}
where
\begin{equation}
\omega=\frac{eB}{m}+\frac{e^{3}B^{3}c^{2}}{A^{2}m^{3}}=\omega_{c}+
\bigtriangleup\omega\,{,}
\end{equation}
and $\bigtriangleup\omega = \omega^3_c c^2/A^2$ is the first correction to the  classical 
frequency $\omega_{c}$.
This is the same correction of Ref. \cite{el} if the
``rigidity'' correction $\beta$-term there obtained is calculated for $v=0$.
Due to the existence of 
higher derivative terms in the Lagrangian, the corrections to the
cyclotron frequency found in \cite{el} depend in fact on the velocity of the 
particle.
 From equation (3.9) we  obtain
\begin{equation}
\frac{\bigtriangleup\omega}{\omega_{c}}=\left(\frac{eBc}{Am}\right)^{2}
\,{.}
\end{equation}
The limit coming from the (g-2)/2 experiment for non relativistic 
electrons trapped in a
magnetic field  $B\sim 50KG$ is \cite{exp}
\begin{equation}
\frac{\bigtriangleup\omega}{\omega_{c}}\leq 2\cdot 10^{-10} \,{.}
\end{equation}
From (3.10) we find
$\bigtriangleup\omega/\omega_{c}\sim 6\cdot 10^{-15}$
when the maximal acceleration is mass-dependent and
$\bigtriangleup\omega/\omega_{c}\sim 2\cdot 10^{-60}$
when the maximal acceleration is considered a universal 
constant. Both results are in agreement with the
experimental upper bound (3.11).

\section{Power Radiated by a Charged Particle}
\setcounter{equation}{0}

The total four-momentum radiated by a charge moving in an electromagnetic 
field is given by the well-known formula \cite{LAN}
\begin{equation}
\Delta P^{\mu}=-\frac{2e^2}{3c}\int |\ddot{x}|^2 dx^{\mu}\,{.}
\end{equation}
If the contributions of the gradients of the electromagnetic fields can be
neglected in (2.5), which is certainly legitimate at the highest gradients
available at present or in a foreseable future, one finds $\Gamma^{\sigma}_{\alpha\beta}\approx 0$,
and (2.4) becomes
\begin{equation}
\ddot{x}^{\sigma}=\frac{e}{mc}\tilde{g}^{\sigma\rho}
F_{\rho\gamma}\dot{x}^{\gamma}\approx\frac{e}{mc}\left(\eta^{\sigma\rho}+
\frac{e^2}{m^2A^2}F^{\sigma}_{\lambda}F^{\lambda\rho}\right)
F_{\rho\gamma}\dot{x}^{\gamma}\,{.}
\end{equation}
Substituting (4.2) into (4.1) we obtain
\begin{equation}
\Delta P^{\mu}=\Delta P^{\mu}_{(0)}+\Delta P^{\mu}_{(A)}\,{,}
\end{equation}
where
\begin{equation}
\Delta P^{\mu}_{(0)}=-\frac{2e^4}{3m^2c^5}\int (F_{\,\,\,\beta}^{\gamma}
\dot{x}^{\beta})
(F_{\gamma\alpha}\dot{x}^{\alpha})dx^{\mu}\,
\end{equation}
is the classical result, and
\begin{equation}
\Delta P^{\mu}_{(A)}=-\frac{2e^4}{3m^2c^5}\frac{e^2}{A^2m^2}
\int (F_{\sigma\gamma}\dot{x}^{\gamma})F^{\sigma}_{\lambda}
F^{\lambda\beta}(F_{\beta\alpha}\dot{x}^{\alpha})dx^{\mu}\,
\end{equation}
represents the maximal acceleration contribution. When $\vec{B}=0$ and
$\vec{E}$ is parallel to $\vec{\beta}=\vec{v}/c$, Eq. (4.5) for
$\mu =0$ becomes
\begin{equation}
\frac{d \varepsilon (A)}{d x^0}=\frac{2e^4}{3m^2c^3}\frac{e^2}{A^2m^2}
|\vec{E}|^4\,{,}
\end{equation}
and (4.4) yields
\begin{equation}
\frac{d \varepsilon}{d x^0}=\frac{2e^4}{3m^2c^3}|\vec{E}|^2\,{.}
\end{equation}
The ratio of the classical dissipation term (4.7) to the maximal acceleration
term (4.6) for electrons is
\begin{equation}
\Gamma=\frac{d\varepsilon/dx^0}{d\varepsilon (A)}=\frac{m^2A^2}{e^2|\vec{E}|^2}
\sim 1.21\,\frac{10^{36}}{|\vec{E}|^2}\,{,}
\end{equation}
while the total dissipated power is
\begin{equation}
P_{tot}=\frac{2e^4}{3m^2c^3}|\vec{E}|^2\left(1+\frac{e^2|\vec{E}|^2}{m^2A^2}
\right)\,{.}
\end{equation}

\section{Conclusions}

Even at their lowest order the maximal acceleration corrections to the 
classical electrodynamics of a particle offer interesting 
aspects. These include the
existence of a critical field beyond which approximation and equations
break down and corrections to the cyclotron frequency of an electron
in a constant magnetic field. While the latter effect is compatible
with present stringent experimental upper limits from the $(g-2)/2$ 
experiment, it may by its very size be difficult to observe directly. 
On the contrary, the field $E_c\sim 2.1\cdot 10^{16} \mbox{N/C}$
for electrons is very close to values $E\sim 5\cdot 10^{15}\mbox{N/C}$
that are presently considered as attainable by focusing $TW$ laser
beams in a plasma \cite{KIM}.
Radiation levels at these values of the electric field also become 
non-negligible as entailed by (4.6), (4.7) and (4.8). In fact the
maximal acceleration contribution increases with the fourth power of 
$|\vec{E}|$ and contributes sizeable amounts to the power radiated at the 
highest field intensities. Increased radiation should of course be
expected on intuitive
grounds. This typical functional increase would constitute the signature
of the process once the promise of high electric fields became a reality.
 
\bigskip

\begin{centerline}
{\bf Acknowledgment}
\end{centerline}

G.P. gladly acknowledges the continued research support of Dr. K. Denford,
Dean of Science, University of Regina.  

\newpage
\begin{centerline}
{\bf REFERENCES}
\end{centerline}

\vfill

\end{document}